\documentclass{svproc}
\usepackage{url}
\usepackage{marvosym}
\usepackage{fancyhdr}
\usepackage{amsmath}
\usepackage{amssymb}
\usepackage{booktabs}
\usepackage{url}
\usepackage{graphicx} 
\usepackage{float} 
\usepackage{subfigure}
\usepackage{xcolor}
\usepackage{colortbl}
\usepackage{multicol}
\usepackage{multirow}
\usepackage{array}
\usepackage{algorithm}
\usepackage{algcompatible}
\usepackage[noend]{algpseudocode}
\usepackage{comment}
\usepackage{enumitem}
\usepackage{caption}

\newcommand{\PreserveBackslash}[1]{\let\temp=\\#1\let\\=\temp}
\newcolumntype{C}[1]{>{\PreserveBackslash\centering}p{#1}}
\newcolumntype{R}[1]{>{\PreserveBackslash\raggedleft}p{#1}}
\newcolumntype{L}[1]{>{\PreserveBackslash\raggedright}p{#1}}

\newcommand{\y}{\mathbf{y}}

\newcommand{\f}{\mathbf{f}} 
\newcommand{\x}{\mathbf{x}}

\begin{document}
\mainmatter              
\title{A Practical and Economical Bayesian Approach to Gas Price Prediction}
\titlerunning{Gaussian Process for Gas Price}  
%
\author{ChihYun Chuang\inst{1} \and TingFang Lee\inst{2}
}
\authorrunning{ChihYun Chuang and TingFang Lee} 
%
\tocauthor{ChihYun Chuang and TingFang Lee}
\institute{AMIS, Taipei, Taiwan\\
\email{chihyun@maicoin.com},\\ 
\and
College of Pharmacy, University of Rhode Island, Rhode Island, USA
\email{tingfanglee@uri.edu}}

\maketitle              

\begin{abstract}
On the Ethereum network, it is challenging to determine a gas price that ensures a transaction will be included in a block within a user's required timeline without overpaying. One way of addressing this problem is through the use of gas price oracles that utilize historical block data to recommend gas prices. However, when transaction volumes increase rapidly, these oracles often underestimate or overestimate the price. In this paper, we demonstrate how Gaussian process models can predict the distribution of the minimum price in an upcoming block when transaction volumes are increasing. This is effective because these processes account for time correlations between blocks. We performed an empirical analysis using the Gaussian process model on historical block data and compared the performance with GasStation-Express and Geth gas price oracles. The results suggest that when transactions volumes fluctuate greatly, the Gaussian process model offers a better estimation. Further, we demonstrated that GasStation-Express and Geth can be improved upon by using a smaller training sample size which \authorrunning{}is properly pre-processed. Based on the results of empirical analysis, we recommended a gas price oracle made up of a hybrid model consisting of both the Gaussian process and GasStation-Express. This oracle provides efficiency, accuracy, and better cost.  
\end{abstract}


\section{Introduction}
On the Ethereum blockchains\cite{Ethereum}, \textit{Gas} refers to the fuel required to conduct a transaction or execute a smart contract\cite{ethyellow}. Since each block on the chain has an upper bound on the amount of gas that can be included in it\footnote{The average block gas limit was around 12,500,000 units of gas at the day of Mar 4, 2021.}, miners maximize profit by prioritizing transactions offering higher gas prices. As with all markets exhibiting supply and demand dynamics, it is advantageous for users to be able to predict and offer the minimum gas price that ensures their transactions will be included in a block within a pre-determined timeline.

Developing such a gas price oracle is complicated by the high variability in transaction volume. Under these conditions, some existing gas price oracles either underestimate the prices needed such that these transactions have to wait a long time to be included, or overestimate the price, which results in users overpaying for the transaction. 

To date, one of the primary methods for gas price prediction is to analyze the pricing structure of pending transactions in large mempools \cite{gasBlockNative}. This method is resource intensive as it requires  accessing large quantities of mempools to obtain enough pending transaction data for analysis. Further, it can only accurately predict the gas prices under the assumption that the data from mempools is correct, something that is difficult for users to verify. Another method is to utilize recent transactions that were included by miners to recommend a price. Some algorithms based on this concept have been used to develop gas price oracles including Geth, EthGasStation, GasStation-Express (abrev. GS-Express) and  the work of Sam M. Werner et al. \cite{Geth,GasStation,GasStationE,ethgas2020,werner2020mingas}. 

One such gas price oracle, GS-Express, proposed using the set of minimum gas prices in the most recent $200$ blocks. The model suggests that the probability of the $\alpha$th percentile of the set is greater than the minimum gas price of the transactions in the next block is $\alpha\%$. In other words, the $\alpha$th percentile of the set has $\alpha\%$ probability to be included in the next block. An additional oracle, Geth, uses the set of minimum gas prices in the most recent $100$ blocks and takes the $60$th percentile of the set as a recommended gas price. These models provide an efficient way to recommend gas prices when the quantity of pending transactions are relatively few. However, when there is a surge of pending transactions, these models will underestimate the prices.

In this paper, we will focus on a novel predictive model that uses recent successful mining blocks to estimate the lowest price a user should offer to obtain a specified probability level that the transaction will be processed. The proposed methodology uses Gaussian process (abrev. GP) models to predict the distribution of the minimum price in the upcoming block. Stochastic processes, including GP, are often used to study numerous stock market micro-structure related questions, including price discovery, competition among related markets, strategic behavior of market participants, and modeling of real time market dynamics. These processes present potentially efficient estimators and predictors for volatility, time-varying correlation structures, trading volume, bid-ask spreads, depth, trading costs, and liquidity risks. The market forces acting within Ethereum markets are very similar to these, making GP an appropriate method to capture the dynamics of Ethereum gas prices. Another attractive feature of stochastic processes is the covariance functions, which allows the model to estimate the time correlation between blocks; that is, it can capture the stronger correlation between closer blocks. Our method provides stable price prediction even when there is a surge in transaction volume. Over the long-term, the gas prices recommended by the model are more economical and practical\footnote{The average time consumed of the GP model is $0.7$ seconds to predict a new price.} compared with existing methods. 

%

Our contributions include the following: 1) We introduce a novel application of Gaussian process to evaluate pricing models and then use it to study the advantages and disadvantages of GS-Express, Geth, and GP. Our findings indicate that GS-Express and Geth over/under-estimated the price when the transaction volume fluctuates greatly, while GP maintained reasonable accuracy. Additionally, GP possesses time efficiencies in model training and prediction. 2) A sensitivity analysis was conducted to study the impact of the training data sizes on the performance of the GS-Express model. This showed reducing the training data size to 50 or 30 can effectively improve prediction. However, this requires a reliable data pre-processing procedure. 3) We propose a practical and economical gas price oracle in Algorithm \ref{alg:gaspriceoracle} which retains the advantages of both GP and GS-Express and avoids the disadvantages of each. Our method is superior in achieving the targeted short-term and long-term success rates among the considered blocks compared with existing methods. Remarkably, except for $P_{50}$ (see Table \ref{tab:mainresult} and \ref{tab:mainresultshortterm}), the average cost of our method is still less than the others.

The outline of this paper is as follows. In section \ref{sec:background}, we introduce the operation of transactions in the Ethereum network and Gaussian processes. In section \ref{sec:Meth}, we establish our methodology, including data pre-processing and modeling. In Section \ref{sec:Analysis} we present the results for GP predictive models and compare with GS-Express and Geth models . Finally in Section \ref{sec:Discussion}, we propose a gas price oracle that is a hybrid of  GP and GS-Express and utilizes the advantages of each. 

\section{Background}\label{sec:background}
In this section, we provide a brief overview of transactions in the Ethereum network and Gaussian processes.

\subsection{Ethereum \& Gas}
Like other permission-less blockchains and cryptocurrencies, Ethereum obtains consensus using a form of cryptographic zero-knowledge proof called  \lq\lq Proof-of-work\lq\lq. In such protocols, a character called \lq\lq miner\rq\rq, groups transactions into a block and appends it to the end of blockchains.  This work is resource consumptive, and thus,  operations using Ethereum require a fee, which is received by miner in exchange for performing the work. Based on the gas price, miners determine which transactions should be included in a block.

A transaction fee is calculated in \textit{Gas}, using a unit called \textbf{wei}$=10^{-18}$ETH or \textbf{Gwei}$=10^{-9}$ETH, where ETH is the currency in Ethereum. The cost of execution is equal to: $$\mbox{gas cost} \times  \mbox{gas price}.$$Here
\begin{itemize}
    \item The gas cost is bounded by the lower bound $21,000$ and the upper bound  \textbf{gas limit}, which represents the maximum amount of gas a user is willing to use for an operation. The precise amount  of  gas cost depends on the complexity of performing \lq\lq smart contracts\rq\rq, which define a set of rules using a Turing-complete programming language. After the transaction is completed, all unused gas is returned to the user's account. If the gas limit is less than the gas cost, then the transaction is viewed as invalid and will be rejected; the gas spent to perform calculations will not be returned to the account. 
    \item The gas price is also determined by the user and represents the price per unit of gas the user is offering to pay. Since a miner's reward is largely determined by the gas price, a higher gas price results in a  greater probability of transactions being selected by miners and grouped into blocks.
\end{itemize}


\subsection{Gaussian Process}

A Gaussian process is a stochastic process that provides a powerful tool for probabilistic inference on distributions over functions. It offers a flexible non-parametric Bayesian framework for estimating latent functions from data. Briefly speaking, Gaussian Process makes prediction with uncertainty. For instance, it will predict that a stock price of the next minute is \$100, with a standard deviation of \$30. Knowing the prediction uncertainty is important for pricing strategies. The rest of this section will follow \cite{rw2006gaussianprocess} to describe the GP regression.

\begin{definition}
A Gaussian process is a collection of random variables, any finite number of which have a joint Gaussian distribution.
\end{definition}

A GP is specified by its mean function and covariance function which determine the functions' smoothness and variability. Given input vectors $\x$ and $\x'$, we define mean function $m(\mathbf{x})$ and the covariance function $k(\mathbf{x}, \mathbf{x}')$ of a real process $f(\mathbf{x})$ as
\begin{align}
m(\mathbf{x})&=E[f(\mathbf{x})]\notag\\
k(\mathbf{x},\mathbf{x}')&=E[(f(\mathbf{x})-m(\mathbf{x}))(f(\mathbf{x}')-m(\mathbf{x}'))]\notag
\end{align}
and will write the Gaussian process as
$$f(\mathbf{x})\sim \mathcal{GP}(m(\mathbf{x}),k(\mathbf{x},\mathbf{x}')).$$ 

Given a training dataset $\mathcal{D}=\{(\x_i, y_i)| i=1, 2, \cdots, n\}$ where $\x$ denotes the input vector and $y$ denotes the target variable. One can consider the Gaussian noise model
$$y_i=f(\x_i)+\mathcal{N}(0, \sigma_n^2).$$
The squared exponential with hyperparameter $\theta=\{\sigma_f, l\}$,
$$k(\x, \x')=\sigma_f^2\exp\big(\frac{-|\x-\x'|^2}{2 l^2}\big),$$ 
is considered the most widely used covariance function. This covariance function is also appropriate in our model, since recent gas prices have stronger correlation.

The joint distribution of the observed target values and the function value at a new input $\x_*$ is
\begin{equation}
    \begin{bmatrix}\y \\ f(\x_*)\end{bmatrix}\sim \mathcal{N}(\mathbf{0}, \begin{bmatrix}K+\sigma_n^2 I & K_*^T \\ K_* & K_{**}\end{bmatrix}),\label{eq:joint_dist}
\end{equation}
where $\y=\begin{bmatrix}y_1 & \cdots y_n\end{bmatrix}^T$, the notation $T$ denotes matrix transportation, $K=[k(\x_i, \x_j)]_{i, j=1, \cdots, n}$ is the covariance matrix, $K_*=[k(\x_*, \x_i)]_{i=1, \cdots, n}$, and $K_{**}=k(\x_*, \x_*)$. Therefore, the posterior predictive distribution is 
\begin{equation}
    f(\x_*)|\mathcal{D}, \x_*, \theta \sim \mathcal{N}(K_*(K+\sigma_n^2 I)^{-1}\y, K_{**}-(K+\sigma_n^2 I)^{-1}K_*^T). \label{eq:pred_dist}
\end{equation}

The accuracy of the GP regression model depends on how well the covariance function is selected. In particular, estimating the hyperparameter of the covariance function is critical to prediction performance. The Laplace approximation framework is often utilized to approximate the predictive posterior distribution and is constructed from the second order Taylor expansion of $\log p(\f|\mathcal{D},\theta)$ around the maximum of the posterior.  It has been shown to provide more precise estimates in much shorter time \cite{laplace2009rue}. 

\section{Methodology} \label{sec:Meth} 
In this section, we explain the steps of data pre-processing. Then, processed data will be fitted into the GP regression model, Geth, and GS-Express. The method to evaluate the performance of each model is introduced in \ref{sec:modelEva}. 

\subsection{Pre-processing}\label{sec:preprocess}
In order to maintain statistical significance, we removed blocks with a number of transactions less than $7$. Further, some blocks have uncommonly low cost transactions. For instance, there are three zero fee transactions in block $11763787$. Such transactions are rare and yet create noise in the models. Therefore, we excluded these transactions by removing  all transactions in which the fees were lower than the 2.5 percentile among all gas prices. The processing steps are as follows:
\begin{itemize}
    \item[1.] Take blocks with more than six transactions.
    \item[2.] Calculate the 2.5 percentile of each block, called $\mathcal{P}_{2.5}$.
    \item[3.] Remove the transactions in which the fees are lower than $\mathcal{P}_{2.5}$. 
    \item[4.] Obtain the minimum gas price in each block, called $y$.
\end{itemize}

\subsection{The Model}
Take $n$ consecutive blocks, $b_1, b_2, \cdots, b_n$, and let the training dataset $\mathcal{D}=\{(i, y_i)|i=1, 2, \cdots, n\}$ where 
$y_i:=\min\{\mbox{ gas prices in block } b_i\}.$
The goal is using the GP regression model to predict $y_{n+1}$ in $b_{n+1}$. 

We consider the Gaussian noise model $\displaystyle y_i=f(i)+\sigma_n^2.$ The squared exponential covariance function is used to estimate the covariance matrix in the joint distribution (\ref{eq:joint_dist}). The posterior predictive distribution (\ref{eq:pred_dist}) is used to predict the mean, $\hat{y}_{n+1}$, and the standard deviation, $\hat{s}_{n+1}$ , of the minimum gas price in the $(n+1)$-th block; that is, $$f(n+1)\sim \mathcal{N}(\hat{y}_{n+1}, \hat{s}_{n+1}^2).$$
More specifically, the above estimation means that the probability of that $\hat{y}_{n+1}$ is greater than the minimum gas price, $y_{n+1}$, of $(n+1)$-th block is 50\%.  We then define that $\hat{y}_{n+1}$ is $P_{50}$ of $(n+1)$-th block. Similarly, $\hat{y}_{n+1}+0.675\cdot\hat{s}_{n+1}$ is $P_{75}$ of $(n+1)$-th block. ($\hat{y}_{n+1}+\hat{s}_{n+1}$ is $P_{84}$\footnote{More precisely, it should be $P_{84.13}$}; $\hat{y}_{n+1}+1.645\cdot\hat{s}_{n+1}$ is $P_{95}$).

\subsection{Model Evaluation}\label{sec:modelEva}
In this section, we will introduce a model comparison criteria, inverse probability weight (IPW), to compare models performance in terms of accuracy and efficiency. Our procedure to compare the GP regression model, GS-Express, and Geth contains the following steps:

\begin{enumerate}
    \item[I.]  Given $n$ consecutive blocks $b_1, b_2, \cdots, b_n$ and $0 <\alpha < 1$, we use the GP regression model, GS-Express, and Geth to predict $P_\alpha$. We then compare the predicted $P_\alpha$ with actual $y_{n+1}$. If $P_\alpha \geq y_{n+1}$, the transaction is viewed as successfully included in block $b_{n+1}$. Define 
    $$T_{n+1}(\alpha)=\begin{cases}
     1, \mbox{ if }P_\alpha\geq y_{n+1} , \\
     0, \mbox{ otherwise.}
    \end{cases}$$
    
    \item[II.] Iterating the model fitting obtains $T_{n+1}, T_{n+2}, \cdots$, and so on. Define the success rate among blocks $b_{s}, b_{s+1}, \cdots, b_{s+t-1}$ as $$R_{s, t, n}(\alpha)=\frac{\sum_{j=0}^{t-1} T_{s+j}(\alpha)}{t}.$$
    Here $n$ is the number of observations of training data.
\end{enumerate}

Note that the function $R_{s,t,n}(\alpha)$ is an increasing function for $0<\alpha<1$.
We use $R_{s, t, n}$ to observe the short term ($t\leq 3n$) and long term ($t\geq 10n$) success rate while using training data with $n$ observations. It is notable that $$T_{n+1}(\alpha)\sim\mbox{Bernoulli}(\alpha)$$ given an ideal gas price oracle. Therefore, in the long run, the success rate of all three methods, GP model, GS-Express, and Geth, using $P_\alpha$ should be approximately $\alpha\%$ (i.e. $R_{s,t,n}$ is a consistent estimator of $\alpha$). That is,
$$\lim_{m\rightarrow \infty}R_{s,m,n}(\alpha)=\alpha\%.$$ When $m$ is not large, $R_{s, m, n}$ can reflect the predictive performance of each method in the short-term. An inefficient gas price oracle can result in a pending transaction, e.g. 50 minutes. Although users can resign\footnote{Use the same nonce and increase the gas price.} the pending transactions from the Ethereum network, a new price is still required from the oracle. Therefore, a gas price oracle needs to perform reasonably in a short period of time.

The ultimate goal is to predict the lowest prices such that the transactions can be included in a block. Higher prices can, of course, result in higher success rates. Therefore, we introduce a measurement, inverse probability weight, which can reflect better pricing strategy.
$$\mbox{IPW}_{s, t, n}(\alpha)=\frac{\mbox{average cost}}{R_{s, t, n}(\alpha)}.$$ 
In words, a small IPW$_{s, t, n}(\alpha)$ represents low cost with high success rate; on the other hand, high cost with low success rate gives large IPW$_{s, t, n}(\alpha)$. We will use this measurement to evaluate model performance in the next section.

\section{Empirical Analysis}\label{sec:Analysis}
We use a laptop with CPU:Intel® Xeon® Processor E3-1505M v5 2.80 GHz and 32gb ram to perform all analysis. Mathematica 12 was utilized for Gaussian process regression model fitting with squared exponential covariance function specified. SageMath was used GS-Express and Geth to operate the model fitting and prediction. All used data can be found in \cite{bayesinagas}.

\subsection{Observations}\label{sec:obser}

The historical block data, block 11753792 to 11823790\footnote{Jan-29-2021 to Feb-09-2021}, were used in the analysis. There were 1450 blocks that have no transaction and 4 blocks have less than 7 transactions. We first removed these blocks and followed steps 2, 3, and 4 in Section \ref{sec:preprocess} to process the remaining blocks. After pre-processing each block, we have 68,545 blocks. We denote those blocks by $b_1, b_2, \cdots, b_{68545}$, e.g. the block number of $b_1$ is 11753792, and $b_{201}$ is 11753994. Taking 200 consecutive blocks as training data, we fit the GP model and GS-Express to the training data. Geth uses only 100 training data. Each model will be used to predict the minimum gas price of the next block. In other words, we fit each of the three models into 1st to 200th blocks, we then predict $P_{50}$, $P_{75}$, $P_{84}$, and $P_{95}$ of the minimum gas price of 201st block. Next, we fit the models into 2nd to 201st blocks, we then predict $P_{50}$, $P_{75}$, $P_{84}$, and $P_{95}$ of the minimum gas price of 202nd block, and so on. The obtained $P_\alpha$ values will be used to compare with the true minimum gas price. Following I. and II. in Section \ref{sec:modelEva}, we will demonstrate the success rate of each model, and compare the performance. 

Ideally, the success rate would be approximately $\alpha\%$ when using $P_\alpha$ to compare with the true minimal price. We find that the long term success rate $R_{201, 68345, 200}$, Table \ref{tab:longsuccess}, of GS-Express and Geth on $P_{50}$ are around 0.5 and GP is 0.36. GS-Express and Geth underestimated the prices on $P_{75}$, $P_{84}$, and $P_{95}$ and GP suggested relatively accurate prices on $P_{75}$, $P_{84}$, and $P_{95}$. The average cost, $$\mbox{the average of predicted price}\cdot 10^{-9},$$ of each method is also reported in Table \ref{tab:longsuccess}.

\begin{table}[h]
  \centering
  \caption{The long term success rate $R_{201, 68545, 200}(\alpha)$ of GP and GS-Express and $R_{201, 68545, 100}(\alpha)$ of Geth. The average cost with $\alpha=50, 75, 84, 95$ of each method is reported at the right hand side of the table. The corresponding inverse probability weights IPW$_{201, 68545, 200}(\alpha)$ are also reported.}
    \begin{tabular}{lC{1cm}C{1cm}C{1cm}C{1cm}C{1cm}C{1cm}C{1cm}C{1cm}}
    \toprule
    \rowcolor[rgb]{ .929,  .929,  .929}      & \multicolumn{4}{c}{Success Rate} & \multicolumn{4}{c}{Average cost (Gwei)} \\
    \rowcolor[rgb]{ .929,  .929,  .929} Method & \multicolumn{1}{c}{$P_{50}$} & \multicolumn{1}{c}{$P_{75}$} & \multicolumn{1}{c}{$P_{84}$} & \multicolumn{1}{c}{$P_{95}$} & \multicolumn{1}{c}{$P_{50}$} & \multicolumn{1}{c}{$P_{75}$} & \multicolumn{1}{c}{$P_{84}$} & \multicolumn{1}{c}{$P_{95}$} \\
    \midrule
    \midrule
    GP   & 0.358 & 0.744 & 0.862 & 0.972 & 128.6 & 168.3 & 187.4 & 225.3 \\
    \rowcolor[rgb]{ .929,  .929,  .929} GS-Express & 0.502 & 0.696 & 0.784 & 0.914 & 141.6 & 168.0  & 178.2 & 197.1 \\
    Geth & 0.500  & 0.712 & 0.798 & 0.922 & 143.1 & 164.7 & 174.1 & 193.3 \\
    \bottomrule
    \end{tabular}%
    
    \begin{tabular}{lC{2.1cm}C{2.1cm}C{2.1cm}C{2.1cm}}
    \toprule
    \rowcolor[rgb]{ .949,  .949,  .949}      & \multicolumn{4}{c}{Inverse Probability Weight} \\
    \rowcolor[rgb]{ .949,  .949,  .949} Method & \multicolumn{1}{c}{$P_{50}$} & \multicolumn{1}{c}{$P_{75}$} & \multicolumn{1}{c}{$P_{84}$} & \multicolumn{1}{c}{$P_{95}$} \\
    \midrule
    \midrule
    GP   & 359.22 & 226.22 & 217.40 & 231.79 \\
    \rowcolor[rgb]{ .949,  .949,  .949} GS-Express & 282.07 & 241.38 & 227.30 & 215.65 \\
    Geth & 286.20 & 231.32 & 218.17 & 209.65 \\
    \bottomrule
    \end{tabular}%

\noindent\footnotesize{Note: The average cost is $\mbox{the average of predicted price}\cdot 10^{-9}$.}
  \label{tab:longsuccess}%
\end{table}

We also calculated the minimum short term success rate $$\min\{ R_{s, m, 200}(\alpha) \mid s=1, 2, \cdots, 68345 \},$$ and reported the minimum success in Table \ref{tab:shortsuccess}. The success rate of consecutive $m=25, 50, 100$ blocks are considered to be fast, average, and slow, respectively, when grouped by miners. From Table \ref{tab:longsuccess} and \ref{tab:shortsuccess}, we observe that GP performs better when using $P_{75}$, $P_{84}$, and $P_{95}$ in a long term and also maintain better success rate in short terms $m=25, 50, 100$.
\begin{table}
\centering
  \caption{The minimum success rate $R_{s, m, 200}(\alpha)$ of GP and GS-Express and $R_{s, m, 100}(\alpha)$ of Geth with $\alpha=50, 75, 84, 95$ of each method. We consider $m=25, 50, 100$ to represent that the transaction is fast, average, and slow to be included in a block.}
    \begin{tabular}{C{1cm}C{0.8cm}C{0.8cm}C{0.8cm}C{0.8cm}C{0.8cm}C{0.8cm}C{0.8cm}C{0.8cm}C{0.8cm}C{0.8cm}C{0.8cm}C{0.8cm}}
    \toprule
     & \multicolumn{4}{c}{\cellcolor[rgb]{ .929,  .929,  .929}GP} & \multicolumn{4}{c}{GS-Express} & \multicolumn{4}{c}{\cellcolor[rgb]{ .929,  .929,  .929}Geth} \\
    $m$ & \cellcolor[rgb]{ .929,  .929,  .929}$P_{50}$ & \cellcolor[rgb]{ .929,  .929,  .929}$P_{75}$ & \cellcolor[rgb]{ .929,  .929,  .929}$P_{84}$ & \cellcolor[rgb]{ .929,  .929,  .929}$P_{95}$ & $P_{50}$ & $P_{75}$ & $P_{84}$ & $P_{95}$ & \cellcolor[rgb]{ .929,  .929,  .929}$P_{50}$ & \cellcolor[rgb]{ .929,  .929,  .929}$P_{75}$ & \cellcolor[rgb]{ .929,  .929,  .929}$P_{84}$ & \cellcolor[rgb]{ .929,  .929,  .929}$P_{95}$ \\
    \midrule
    \midrule
    25    & \cellcolor[rgb]{ .929,  .929,  .929}0 & \cellcolor[rgb]{ .929,  .929,  .929}0.12 & \cellcolor[rgb]{ .929,  .929,  .929}0.24 & \cellcolor[rgb]{ .929,  .929,  .929}0.36 & 0     & 0     & 0     & 0.12  & \cellcolor[rgb]{ .929,  .929,  .929}0 & \cellcolor[rgb]{ .929,  .929,  .929}0 & \cellcolor[rgb]{ .929,  .929,  .929}0 & \cellcolor[rgb]{ .929,  .929,  .929}0.24 \\
    50    & \cellcolor[rgb]{ .929,  .929,  .929}0.04 & \cellcolor[rgb]{ .929,  .929,  .929}0.20 & \cellcolor[rgb]{ .929,  .929,  .929}0.32 & \cellcolor[rgb]{ .929,  .929,  .929}0.54 & 0.02  & 0.06  & 0.06  & 0.18  & \cellcolor[rgb]{ .929,  .929,  .929}0.04 & \cellcolor[rgb]{ .929,  .929,  .929}0.06 & \cellcolor[rgb]{ .929,  .929,  .929}0.08 & \cellcolor[rgb]{ .929,  .929,  .929}0.40 \\
    100   & \cellcolor[rgb]{ .929,  .929,  .929}0.09 & \cellcolor[rgb]{ .929,  .929,  .929}0.33 & \cellcolor[rgb]{ .929,  .929,  .929}0.42 & \cellcolor[rgb]{ .929,  .929,  .929}0.71 & 0.07  & 0.12  & 0.16  & 0.32  & \cellcolor[rgb]{ .929,  .929,  .929}0.10 & \cellcolor[rgb]{ .929,  .929,  .929}0.19 & \cellcolor[rgb]{ .929,  .929,  .929}0.23 & \cellcolor[rgb]{ .929,  .929,  .929}0.50 \\
    \bottomrule
    \end{tabular}%
  \label{tab:shortsuccess}%
\end{table}

We now focus on the success rate of $P_{75}$ of each method. GP gave 0.744 success rate while GS-Express and Geth had 0.696 and 0.712. We derived an $\alpha$ such that $P_\alpha$ from GP provided a comparable level of success rate, but lower cost than GS-Express and Geth, see Table \ref{tab:midvalue}. Therefore, in the long term, GP has the advantage in cost and success rate. However, GS-Express can have better performance when using less training data.
\begin{table}
\centering
  \caption{The minimum short term success rate  $R_{s, m, 200}$, long term success rate $R_{201, 68545, 200}$, and the average cost using GP predicted $P_{72.24}$.}
    \begin{tabular}{ccC{2cm}r}
    \toprule
    \multicolumn{2}{c}{Short term success rate} & \multicolumn{2}{c}{Long term} \\
    \midrule
    \midrule
    $m=25$ & 0.12 & \multicolumn{1}{l}{Success rate} & 0.712 \\
    $m=50$ & 0.18 & \multicolumn{1}{l}{Average cost} & \multicolumn{1}{l}{164.16 Gwei} \\
    $m=100$ & 0.30  &  \multicolumn{1}{l}{IPW}  & 230.56  \\
    \bottomrule
    \end{tabular}%
  \label{tab:midvalue}%

\end{table}%

\subsection{More Observations}\label{sec:moreobserv}
The gas prices showed large fluctuations during block 11903793 to 11917694\footnote{Feb-22-2021 to Feb-24-2021}. We also provided analogous results for these blocks transaction data. According to the previous observations, we reduced the training data points for GS-Express to 50 or 30. After pre-processing these recent block data, we have 13627 blocks denoted by $b_1, b_2, \cdots, b_{13627}$, e.g. the block number of $b_1$ is 11903793, and $b_{201}$ is 11903999. The long term success rate $R_{201, 13627, 200}$ of GP and $R_{201, 13627, 50}$ and $R_{201, 13627, 30}$ of GS-Express, and the average costs are reported in Table \ref{tab:longsuccessV2}. The minimum short term success rates ($m=25, 50, 100$) of each model can be found in Table \ref{tab:shortsuccessV2}. The results are consistent with the observations in Section \ref{sec:obser}.

\begin{table}[h]
  \centering
  \caption{The long term success rate $R_{201, 13627, 200}(\alpha)$ of GP prediction, $R_{201, 13427, 30}(\alpha)$ and $R_{201, 13427, 50}(\alpha)$ using GS-Express.  And the average cost with $\alpha=50, 75, 84, 95$ of each method.  The corresponding inverse probability weights IPW$_{201, 68545, 200}(\alpha)$ are also reported.}
    \begin{tabular}{lC{1cm}C{1cm}C{1cm}C{1cm}C{1cm}C{1cm}C{1cm}C{1cm}}
    \toprule
    \rowcolor[rgb]{ .929,  .929,  .929}      & \multicolumn{4}{c}{Success Rate} & \multicolumn{4}{c}{Average cost (Gwei)} \\
    \rowcolor[rgb]{ .929,  .929,  .929} Method & \multicolumn{1}{c}{$P_{50}$} & \multicolumn{1}{c}{$P_{75}$} & \multicolumn{1}{c}{$P_{84}$} & \multicolumn{1}{c}{$P_{95}$} & \multicolumn{1}{c}{$P_{50}$} & \multicolumn{1}{c}{$P_{75}$} & \multicolumn{1}{c}{$P_{84}$} & \multicolumn{1}{c}{$P_{95}$} \\
    \midrule
    \midrule
    GP   & 0.375 & 0.79 & 0.911 & 0.982 & 248.5 & 314.5 & 346.3 & 409.4 \\
    \rowcolor[rgb]{ .929,  .929,  .929} GS-Express (30) & 0.516 & 0.70  & 0.786 & 0.882 & 270.9 & 297.0  & 312.4 & 331.8 \\
    GS-Express (50) & 0.506 & 0.70  & 0.786 & 0.893 & 270.2 & 301.1 & 317.5 & 341.0 \\
    \bottomrule
    \end{tabular}%
    
    \begin{tabular}{lC{2.1cm}C{2.1cm}C{2.1cm}C{2.1cm}}
    \toprule
    \rowcolor[rgb]{ .949,  .949,  .949}  & \multicolumn{4}{c}{Inverse Probability Weight} \\
    \rowcolor[rgb]{ .949,  .949,  .949} Method & \multicolumn{1}{c}{$P_{50}$} & \multicolumn{1}{c}{$P_{75}$} & \multicolumn{1}{c}{$P_{84}$} & \multicolumn{1}{c}{$P_{95}$} \\
    \midrule
    \midrule
    GP   & 662.67 & 398.10 & 380.13 & 416.60 \\
    \rowcolor[rgb]{ .949,  .949,  .949} GS-Express (30) & 525.00  & 424.29 & 397.46 & 376.19 \\
    GS-Express (50) & 533.99 & 430.14 & 403.94 & 381.86 \\
    \bottomrule
    \end{tabular}%
  \label{tab:longsuccessV2}%

\end{table}

\begin{table}[h]
\centering
  \caption{The minimum success rate $R_{s, m}(\alpha)$ with $\alpha=50, 75, 84, 95$ of each method. We consider $m=25, 50, 100$ to represent that the transaction is fast, average, and slow to be included in block.}
    \begin{tabular}{C{1cm}C{0.8cm}C{0.8cm}C{0.8cm}C{0.8cm}C{0.8cm}C{0.8cm}C{0.8cm}C{0.8cm}C{0.8cm}C{0.8cm}C{0.8cm}C{0.8cm}}
    \toprule
     & \multicolumn{4}{c}{\cellcolor[rgb]{ .929,  .929,  .929}GP} & \multicolumn{4}{c}{GS-Express(30)} & \multicolumn{4}{c}{\cellcolor[rgb]{ .929,  .929,  .929}GS-Express(50)} \\
    $m$ & \cellcolor[rgb]{ .929,  .929,  .929}$P_{50}$ & \cellcolor[rgb]{ .929,  .929,  .929}$P_{75}$ & \cellcolor[rgb]{ .929,  .929,  .929}$P_{84}$ & \cellcolor[rgb]{ .929,  .929,  .929}$P_{95}$ & $P_{50}$ & $P_{75}$ & $P_{84}$ & $P_{95}$ & \cellcolor[rgb]{ .929,  .929,  .929}$P_{50}$ & \cellcolor[rgb]{ .929,  .929,  .929}$P_{75}$ & \cellcolor[rgb]{ .929,  .929,  .929}$P_{84}$ & \cellcolor[rgb]{ .929,  .929,  .929}$P_{95}$ \\
    \midrule
    \midrule
    25    & \cellcolor[rgb]{ .929,  .929,  .929}0 & \cellcolor[rgb]{ .929,  .929,  .929}0.08 & \cellcolor[rgb]{ .929,  .929,  .929}0.24 & \cellcolor[rgb]{ .929,  .929,  .929}0.32 & 0     & 0     & 0     & 0.28  & \cellcolor[rgb]{ .929,  .929,  .929}0 & \cellcolor[rgb]{ .929,  .929,  .929}0 & \cellcolor[rgb]{ .929,  .929,  .929}0 & \cellcolor[rgb]{ .929,  .929,  .929}0.16 \\
    50    & \cellcolor[rgb]{ .929,  .929,  .929}0.04 & \cellcolor[rgb]{ .929,  .929,  .929}0.24 & \cellcolor[rgb]{ .929,  .929,  .929}0.38 & \cellcolor[rgb]{ .929,  .929,  .929}0.52 & 0.08  & 0.10  & 0.14  & 0.42  & \cellcolor[rgb]{ .929,  .929,  .929}0.06 & \cellcolor[rgb]{ .929,  .929,  .929}0.08 & \cellcolor[rgb]{ .929,  .929,  .929}0.10 & \cellcolor[rgb]{ .929,  .929,  .929}0.28 \\
    100   & \cellcolor[rgb]{ .929,  .929,  .929}0.09 & \cellcolor[rgb]{ .929,  .929,  .929}0.30 & \cellcolor[rgb]{ .929,  .929,  .929}0.42 & \cellcolor[rgb]{ .929,  .929,  .929}0.63 & 0.12  & 0.16  & 0.21  & 0.52  & \cellcolor[rgb]{ .929,  .929,  .929}0.12 & \cellcolor[rgb]{ .929,  .929,  .929}0.12 & \cellcolor[rgb]{ .929,  .929,  .929}0.16 & \cellcolor[rgb]{ .929,  .929,  .929}0.39 \\
    \bottomrule
    \end{tabular}%
  \label{tab:shortsuccessV2}%

\end{table}

Similarly, we find that the prediction $P_{69.67}$ from GP gave a comparable level of success rate with the other methods, but lower cost, see Table \ref{tab:midvalueV2}.

\begin{table}[h]
\centering
  \caption{The minimum short term success rate  $R_{s, m, 200}$, long term success rate $R$, and the average cost using GP predicted $P_{69.67}$.}
    \begin{tabular}{ccC{2cm}r}
    \toprule
    \multicolumn{2}{c}{Short term success rate} & \multicolumn{2}{c}{Long term} \\
    \midrule
    \midrule
    $m=25$ & 0.08 & \multicolumn{1}{l}{Success rate} & 0.701 \\
    $m=50$ & 0.20  & \multicolumn{1}{l}{Average cost} & \multicolumn{1}{l}{298.85 Gwei} \\
    $m=100$ & 0.26 &  \multicolumn{1}{l}{IPW}    & 426.32 \\
    \bottomrule
    \end{tabular}%
  \label{tab:midvalueV2}%

\end{table}%

\section{Discussion} \label{sec:Discussion}
The goal of our study is to develop a gas price oracle which ensures that transactions will be included in a block within a user required timeline without overpaying. The proposed the GP regression provided an efficient prediction of gas prices, especially when the transaction volumes increase rapidly. 

\begin{figure}
\centering
\caption{The comparison of the true minimum gas prices in each block  and the prediction of $\mbox{P}_{75}$ from each method. Block number $11799554$ to $11800236$ (top) and block number $11907262$ to $11907569$ (bottom) are shown here.}
\subfigure[block number:11799554-11800236]{\includegraphics[scale=0.17]
{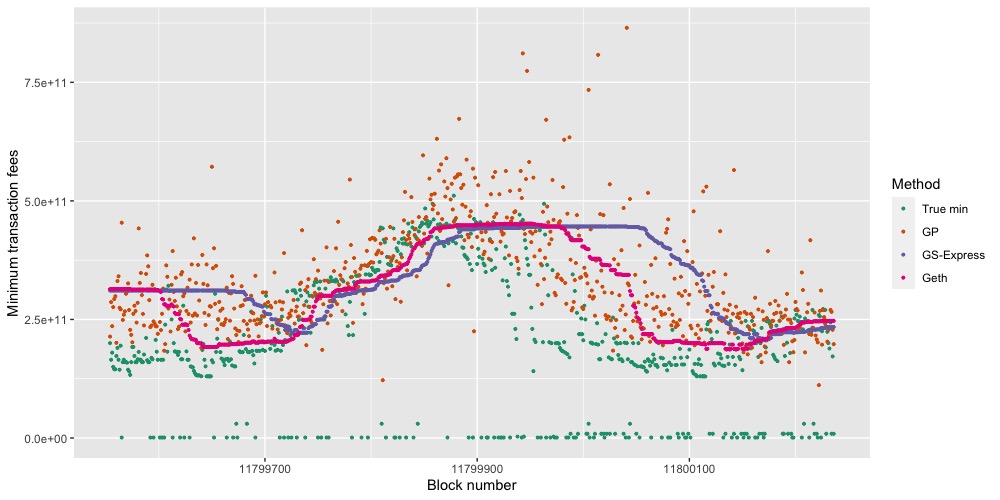}}\subfigure[block number:11907262-11907569]{\includegraphics[scale=0.17]{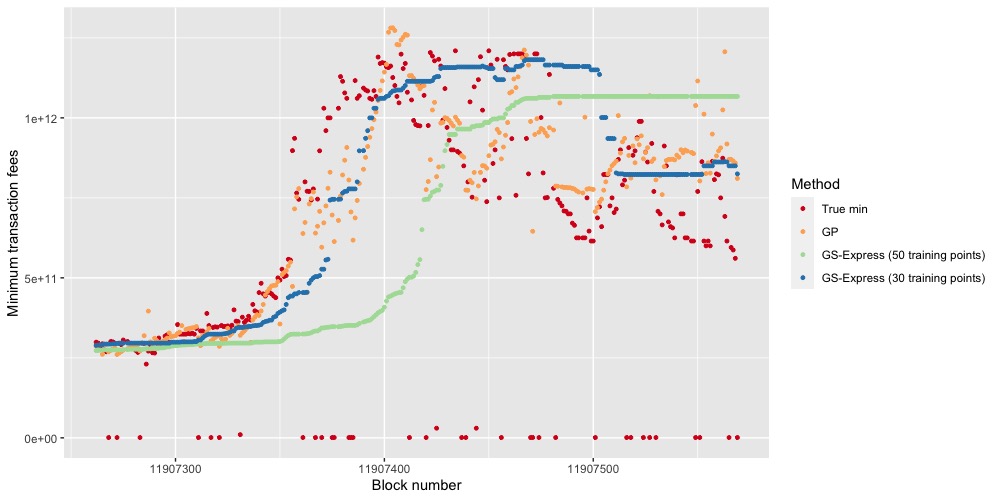}}    
\label{Fig.observ}

\end{figure}

When using various amounts of data, e.g., 200, 100, 50, or 30, the GS-Express model performed poorly from block $11799554$ to $11799759$ and $11907262$ to $11907569$. Figure \ref{Fig.observ} compared the prediction of $P_{75}$ from each method to the true minimum gas price in each block. When gas prices increased, GS-Express often underestimated the price and resulted in a pending transaction until the price stabilized or dropped. Furthermore, GS-Express overestimated the price when the price just dropped from a peak, which results in the user overpaying.

When the pending transaction volume remains stable, GS-Express performs better if we reduce the training data observations (from 200 down to 50 or 30). Reducing the training data can be a risk due to abnormal transactions such as zero fee transactions which may create more noise for the models. Pre-processing the data can effectively reduce the impact of these abnormal data points. Therefore, our observations from the empirical analysis are as follows:
\begin{itemize}
    \item[1.] GP maintains a better success rate with little overpayment when transaction volumes are increasing rapidly.
    \item[2.] The prediction of Geth and GS-Express can be improved by reducing training data when the transaction volume fluctuate greatly. However, abnormal transactions can interfere with the models. Therefore, pre-processing data is an important step.
    \item[3.] Long term success rates of all 3 methods are comparable. However, Geth and GS-Express often underestimate the gas price when gas price rise rapidly. 
\end{itemize}

In addition to using GP only, we propose the following gas price oracle, Algorithm \ref{alg:gaspriceoracle}, which consists of GP and GS-Express and depends on the change of instant success rates. Instant success rate $R$ can be used to monitor the bias of the GP and GS-Express estimators. When the gas prices are stable, GS-Express with a small training sample size performs well. When gas prices increases rapidly, the success rate $R$ is smaller than the expected value $\alpha$, and users can switch to GP to maintain the expected $\alpha$. If $R$ is higher than $\alpha$, the value $\alpha$ should be adjusted to a lower level. This oracle offers efficiency, success rate, and better cost.

\begin{algorithm}
    \caption{Gas Price Oracle}\label{alg:gaspriceoracle}

\hspace*{\algorithmicindent} \textbf{Input: the desired success rate $\alpha$, $n_{\mbox{\tiny GS}}$(resp. $n_{\mbox{\tiny GP}}$) the size of training data of GS-Express(resp. GP), and an allowed error $e$ .} \\
\hspace*{\algorithmicindent} \textbf{Output: a prediction of the block with the number $s+ n_{\mbox{\tiny GS}}$.} 

\begin{algorithmic}[1]
\State Perform pre-process:
\begin{itemize}
    \item[i.] Take blocks with more than six transactions.
    \item[ii.] Calculate the 2.5 percentile of each block, called $P_{2.5}$
    \item[iii.] Remove the transactions in which the fees are lower than $P_{2.5}$.
    \item[iv.] Construct a set $M$ of the minimum gas price in each block.
\end{itemize}
\State Let the, $P_{\alpha}$, prediction of GP(resp. GS-Express) be $P_{\mbox{\tiny GP},\alpha}$(resp. $P_{\mbox{\tiny GS}, \alpha}$). With $s$ advancing, the success rate $R_{s, n_{\mbox{\tiny GS}}, n_{\mbox{\tiny GS}}}(\alpha)$ of GS-Express keeps updating, do:  
\begin{itemize}
    \item[a.] the case $R_{s, n_{\mbox{\tiny GS}},n_{\mbox{\tiny GS}}}(\alpha) < \alpha - e$:  The output is $\max\{P_{\mbox{\tiny GP},\alpha}, P_{\mbox{\tiny GS},\alpha}\}$.
    \item[b.] the case $\alpha-e\leq R_{s, n_{\mbox{\tiny GS}},n_{\mbox{\tiny GS}}}(\alpha) \leq \alpha + e$: The output is $P_{\mbox{\tiny GS},\alpha}$.
    \item[c.] the case $R_{s, n_{\mbox{\tiny GS}},n_{\mbox{\tiny GS}}}(\alpha) > \alpha + e$: Use intermediate value theorem to find $\alpha'$ such that $\alpha-e\leq R_{s, n_{\mbox{\tiny GS}}, n_{\mbox{\tiny GS}}}(\alpha') \leq \alpha + e$. If such $\alpha'$  does not exist, one takes $\alpha' =\alpha$. Then the output is $P_{\mbox{\tiny GS},\alpha'}$.
\end{itemize}
\end{algorithmic}
\end{algorithm}

\begin{table}[H]
 \centering
  \caption{The long term success rate, the average cost, and the inverse probability weights with $\alpha=50, 75, 84, 95$ of our method, GS-Express, and Geth using data from block 11753792 to 11823790 (top two tables). The bottom two tables report the long term success rate, the average cost, and the inverse probability weights using data from block 11903793 to 11917694.}
  
    \begin{tabular}{lC{1cm}C{1cm}C{1cm}C{1cm}C{1cm}C{1cm}C{1cm}C{1cm}}
    \toprule
    \rowcolor[rgb]{ .929,  .929,  .929}      & \multicolumn{4}{c}{Success Rate} & \multicolumn{4}{c}{Average cost (Gwei)} \\
    \rowcolor[rgb]{ .929,  .929,  .929} Method & \multicolumn{1}{c}{$P_{50}$} & \multicolumn{1}{c}{$P_{75}$} & \multicolumn{1}{c}{$P_{84}$} & \multicolumn{1}{c}{$P_{95}$} & \multicolumn{1}{c}{$P_{50}$} & \multicolumn{1}{c}{$P_{75}$} & \multicolumn{1}{c}{$P_{84}$} & \multicolumn{1}{c}{$P_{95}$} \\
    \midrule
    \midrule
    Our method   & 0.52 & 0.73 & 0.81 & 0.92 & 143.0 & 158.4 & 168.1 & 189.7 \\
    \rowcolor[rgb]{ .929,  .929,  .929} GS-Express & 0.50 & 0.70 & 0.78 & 0.91 & 141.6 & 168.0  & 178.2 & 197.1 \\
    Geth & 0.50  & 0.71 & 0.80 & 0.92 & 143.1 & 164.7 & 174.1 & 193.3 \\
    \bottomrule
    \end{tabular}%

    \begin{tabular}{lC{2.1cm}C{2.1cm}C{2.1cm}C{2.1cm}}
    \toprule
    \rowcolor[rgb]{ .949,  .949,  .949}      & \multicolumn{4}{c}{Inverse Probability Weight} \\
    \rowcolor[rgb]{ .949,  .949,  .949} Method & \multicolumn{1}{c}{$P_{50}$} & \multicolumn{1}{c}{$P_{75}$} & \multicolumn{1}{c}{$P_{84}$} & \multicolumn{1}{c}{$P_{95}$} \\
    \midrule
    \midrule
    Our method & 275.00  & 216.99 & 207.53 & 206.20 \\
    \rowcolor[rgb]{ .949,  .949,  .949} GS-Express & 283.20 & 240.00  & 228.46  & 216.59  \\
    Geth & 286.20 & 231.97  & 217.63 & 210.11 \\
    \bottomrule
    \end{tabular}%
    
\begin{tabular}{lC{1cm}C{1cm}C{1cm}C{1cm}C{1cm}C{1cm}C{1cm}C{1cm}}
    \toprule
    \rowcolor[rgb]{ .929,  .929,  .929}      & \multicolumn{4}{c}{Success Rate} & \multicolumn{4}{c}{Average cost (Gwei)} \\
    \rowcolor[rgb]{ .929,  .929,  .929} Method & \multicolumn{1}{c}{$P_{50}$} & \multicolumn{1}{c}{$P_{75}$} & \multicolumn{1}{c}{$P_{84}$} & \multicolumn{1}{c}{$P_{95}$} & \multicolumn{1}{c}{$P_{50}$} & \multicolumn{1}{c}{$P_{75}$} & \multicolumn{1}{c}{$P_{84}$} & \multicolumn{1}{c}{$P_{95}$} \\
    \midrule
    \midrule
    Our method   & 0.52 & 0.75 & 0.842 & 0.92 & 272.7 & 303.6 & 323.5 & 352.3 \\
    \rowcolor[rgb]{ .929,  .929,  .929} GS-Express  & 0.52 & 0.70  & 0.79 & 0.90 & 264.8 & 316.2 & 335.2 & 367.8 \\
    Geth & 0.50 & 0.70  & 0.78 & 0.91 & 268.8 & 309.4  & 324.5 & 358.7 \\
    \bottomrule
    \end{tabular}%

    \begin{tabular}{lC{2.1cm}C{2.1cm}C{2.1cm}C{2.1cm}}
    \toprule
    \rowcolor[rgb]{ .949,  .949,  .949}      & \multicolumn{4}{c}{Inverse Probability Weight} \\
    \rowcolor[rgb]{ .949,  .949,  .949} Method & \multicolumn{1}{c}{$P_{50}$} & \multicolumn{1}{c}{$P_{75}$} & \multicolumn{1}{c}{$P_{84}$} & \multicolumn{1}{c}{$P_{95}$} \\
    \midrule
    \midrule
    Our method & 524.42  & 404.80 & 384.20 & 382.93  \\
    \rowcolor[rgb]{ .949,  .949,  .949} GS-Express & 509.23  & 451.71  & 424.30  & 408.67 \\
    Geth & 537.60 & 442.00  & 416.03 & 394.18 \\
    \bottomrule
    \end{tabular}%

\noindent\footnotesize{Note: The average cost is $\mbox{the average of predicted price}\cdot 10^{-9}$.}
  \label{tab:mainresult}%

\end{table}

To evaluate the proposed gas price oracle, we use the gas prices in block 11799554 to 11800236 and 11903999 to 11917694. The gas prices in block 11799554 to 11800236 changed gradually and in 11903999 to 11917694 changed rapidly. We take $n_{\mbox{\tiny GS}} =30$, $n_{\mbox{\tiny GP}}=200$, and $e=0.1$ in our oracle. Our gas price oracle has smaller inverse probability weighting under most circumstances. The results suggested that our oracle has lower costs with higher long term success rates while also preserving the desired short term success rates, see Table \ref{tab:mainresult} and \ref{tab:mainresultshortterm}. 

Potential future work includes practicing the proposed gas price prediction procedure in real-time and verifying the efficiency and accuracy of this procedure. This will also enhance the study of short term success rate and real waiting times. Additionally, different covariance functions of the GP regression models should also be studied closely to determine which covariance function would be the best suited for predicting gas prices. 

\begin{table}
\centering
  \caption{The short term success rate of each method with $m=25, 50, 100$ from the data of blocks 11753792 to 11823790  and 11903793 to 11917694.}
    \begin{tabular}{C{1cm}C{1cm}C{1cm}C{1cm}C{1cm}C{1cm}C{1cm}C{1cm}C{1cm}C{1cm}}
    \toprule
         & \multicolumn{9}{c}{block 11753792 to 11823790} \\
         & \multicolumn{3}{c}{Our method} & \multicolumn{3}{c}{GS-Express} & \multicolumn{3}{c}{Geth} \\
    $m$    & 25   & \multicolumn{1}{c}{50} & \multicolumn{1}{c}{100} & 25   & 50   & 100  & 25   & 50   & 100 \\
    \midrule
    \midrule
    \rowcolor[rgb]{ .929,  .929,  .929} $P_{50}$  & \multicolumn{1}{c}{0.04} & \multicolumn{1}{c}{0.08} & \multicolumn{1}{c}{0.18} & 0    & 0.02 & 0.07 & 0    & 0.04 & 0.10 \\
    $P_{75}$  & 0.12 & \multicolumn{1}{c}{0.22} & \multicolumn{1}{c}{0.34} & 0    & 0.06 & 0.12 & 0    & 0.06 & 0.19 \\
    \rowcolor[rgb]{ .929,  .929,  .929} $P_{84}$  & 0.20 & \multicolumn{1}{c}{0.38} & \multicolumn{1}{c}{0.50} & 0    & 0.06 & 0.16 & 0    & 0.08 & 0.23 \\
    $P_{95}$  & \multicolumn{1}{c}{0.68} & \multicolumn{1}{c}{0.72} & 0.81 & 0.12 & 0.18 & 0.32 & 0.28 & 0.40  & 0.50 \\
    \midrule

    \midrule
         & \multicolumn{9}{c}{block 11903793 to 11917694} \\
         & \multicolumn{3}{c}{Our method} & \multicolumn{3}{c}{GS-Express} & \multicolumn{3}{c}{Geth} \\
    $m$    & 25   & \multicolumn{1}{c}{50} & \multicolumn{1}{c}{100} & 25   & 50   & 100  & 25   & 50   & 100 \\
    \midrule
    \midrule
    \rowcolor[rgb]{ .929,  .929,  .929} $P_{50}$  & 0    & \multicolumn{1}{c}{0.08} & \multicolumn{1}{c}{0.17} & 0    & 0.04 & 0.06 & 0    & 0.04 & 0.08 \\
    $P_{75}$  & 0.12 & \multicolumn{1}{c}{0.28} & \multicolumn{1}{c}{0.35} & 0    & 0.04 & 0.10  & 0    & 0.06 & 0.12 \\
    \rowcolor[rgb]{ .929,  .929,  .929} $P_{84}$  & 0.32 & \multicolumn{1}{c}{0.46} & \multicolumn{1}{c}{0.50} & 0    & 0.06 & 0.12 & 0    & 0.08 & 0.12 \\
    $P_{95}$  & 0.52 & 0.64 & 0.73 & 0    & 0.08 & 0.13 & 0.08 & 0.14 & 0.22 \\
    \bottomrule
    \end{tabular}%
  \label{tab:mainresultshortterm}%

\end{table}%

\section*{Acknowledgment} 
We would like to thank all of the AMIS data management teams, and participants who contributed to this project. We also thank Yu-Te Lin and Gavino Puggioni for their useful comments and feedbacks in earlier drafts and all the reviewers' helpful comments and suggestions. 

\bibliographystyle{splncs03_unsrt}
\bibliography{bib}

\end{document}